\documentclass[%
 reprint,
nofootinbib,
 amsmath,amssymb,
 aps,
]{revtex4-1}

\def\be{\begin{equation}}
\def\ee{\end{equation}}
\def\bi{\begin{itemize}}
\def\ei{\end{itemize}}
\def\ben{\begin{enumerate}}
\def\een{\end{enumerate}}
\def\bea{\begin{eqnarray}}
\def\eea{\end{eqnarray}}

\usepackage{color}
\usepackage{graphicx}
\usepackage{dcolumn}
\usepackage{bm}
\usepackage{natbib}
\usepackage{hyperref}

\begin{document}

\preprint{APS/123-QED}

\title{Saturating The Bekenstein-Hawking Entropy Bound With Initial Data Sets For Gravitational Collapse }

\author{Sina Bahrami}
 \email{sb933@cornell.edu}
\affiliation{Department of Physics, Cornell University 
}%

\date{\today}

\begin{abstract}
It is possible to find initial states for gravitational collapse whose
entropy approximately saturates the Bekenstein-Hawking entropy of the final black hole.
The prototypical example of such a state is that envisaged by Zurek
and Thorne, and also
by Susskind: for a black hole of mass $M$, a number $\sim M^2$ of
quanta with energies of order $\sim M^{-1}$ are accreted on a
timescale of order $\sim M^3$, an approximate time-reverse of Hawking evaporation.
There is lore that all initial states which saturate the
Bekenstein-Hawking entropy must involve a formation timescale of this
order, $\sim M^3$, and not the much shorter dynamical timescale $\sim
M$.
Counterexamples to this lore have been found by Sorkin, Wald and
Zhang, and also by Hsu and Reeb, in the form of semiclassical initial
data sets.  However the spacetimes that
correspond to these counterexamples possess white holes in the past, as
well as black holes in the future, which casts doubt on their physical relevance.
We modify the counterexamples to eliminate the white holes, yielding formation timescales of order $\sim M^2$, 
and argue that the lore is unfounded.

\end{abstract}

\pacs{Valid PACS appear here}
\maketitle


\section{\label{intro} Introduction}

Forty some years have passed since Bekenstein conjectured
that a black hole is a thermal system with
an entropy proportional
to the area of its event horizon \citep{beken}.
Subsequent careful analyses by Hawking confirmed the
thermal nature of black holes \citep{hawk1,hawk2}. Since then,
the microphysical interpretation of the Bekenstein-Hawking entropy
has been the subject of much speculation and discussion \cite{Strominger:1997eq,Ashtekar:1997yu,waldrev}.
One of the earliest proposals was that black hole
entropy counts
the number of distinct initial data sets from which the black hole
might have been formed \citep{beken, hawk1}.
The motivation for this proposal is that
the number of states of a system is not
expected to change if the system undergoes a unitary quantum mechanical evolution.
Thus, one expects that
the number $N$ of distinct initial data sets that can collapse to form
a given black hole must match its number of internal states, and
correspondingly match the Bekenstein-Hawking entropy, $\ln N \sim
M^2$, where $M$ is the black hole mass \footnote{Throughout this paper we work in the units of $k=c=\hbar=G=1$.}.

As is well known, for black holes that are formed in astrophysical
collapses, the entropies $S$ of the pre-collapse configurations do not
saturate the Bekenstein-Hawking value.  When gravity is weak, the
maximum entropy that can be contained in a region of size $R$ using an
energy $E$ is that of thermal radiation, $S \sim (E R)^{3/4}$, and for
the final stages of collapse taking $E \sim M$ and $R \sim M$ gives $S
\sim M^{3/2}$, far smaller than the Bekenstein-Hawking value.
Although this estimate is based on the assumption of weak gravity
which is violated during the late stages of collapse, one expects it
to be a robust upper bound for systems which start in a weak gravity regime.
The final stages of such collapses occur on a timescale $\sim M$.

A different formation scenario can saturate\footnote{In the weak sense that
  the logarithm of the number of states scales like $M^2$, not in the
  stronger sense that it is $M^2/4$ plus subleading corrections.} the Bekenstein-Hawking
entropy.  As first pointed out by Zurek and Thorne \cite{thorne} (see
also \cite{harlow})
small black holes of initial mass $m$ much smaller than the final mass $M$ can
accrete quanta of energies $E \sim 1/m$ one-by-one, each taking a time
of order $m$.  The black hole then grows in an approximate time
reverse of the Hawking evaporation process, yielding a final mass $M$
after accreting $ \sim M^2$ quanta.  The number $N$ of such initial
states is then given by $\ln N \sim S \sim M^2$.
However, this formation process occurs on a much longer timescale than
astrophysical collapse, of order $\tau \sim M^3$
\footnote{Also see \cite{dvali}, which shares some similarities 
with the Thorne Zurek construction.}.

One can also construct scenarios which are intermediate between the
two extremes of dynamical collapse and the Thorne-Zurek scenario.
We consider a black hole accreting quanta of a quantum field where
each sector $l,m$ for $l \le \l_{\rm max}$ is an independent $1+1$
dimensional thermal gas with quanta of energy $E \sim l_{\rm max}/M$
and size $\tau$.  The total number of quanta per sector is $\tau E
\sim \tau l_{\rm max}/M$, and the total number of quanta is given by
multiplying by the number $l_{\rm max}^2$ of sectors, giving
$S \sim \tau l_{\rm max}^3/M$.  Similarly the total energy per sector
is $\tau E^2 \sim \tau l_{\rm max}^2/M^2$, and the total energy is
then $\tau l_{\rm max}^4/M^2$.  Equating this to $M$ and eliminating
$l_{\rm max}$ yields a relation between entropy $S$, black hole mass
$M$ and formation timescale which interpolates between the two extreme
examples:
\be
S \sim M^{5/4} \tau^{1/4}.
\label{genericrel}
\ee

The examples discussed above, and specifically the
relation \eqref{genericrel}, all support a lore in the community that
generic black hole microstates must be formed by collapse processes on
timescales much longer than $M$.
In particular, one might conjecture that the relation
\eqref{genericrel} is generic, valid in order of magnitudes for all
collapse scenarios.
The purpose of this paper is to
investigate this lore and see if there is any evidence against it.

There have been a number of attempts to challenge
the conjectured relation \eqref{genericrel}.
Most famous among these attempts are the initial data sets
first constructed by Sorkin \emph{et.al.} in \citep{sorkin}
which later became known as "monsters"\citep{hsu1} \footnote{Also see \citep{sorkin,hsu2} for the so-called bags of gold
initial data sets. Similar to monsters, the full Cauchy development of these
initial data sets involve both a white hole and a black hole singularity.}.
A monster is a time symmetric and asymptotically flat spacelike Cauchy surface
that is foliated by nearly marginally trapped 2-spheres. The stress energy tensor is everywhere vacuum except for a compact spherical region which is filled with a thermalized gas of radiation. Monsters are expected to form black holes within a timescale on the order of $M$. A large entropy on the order of $M^2$
is achieved for monsters due to the enhanced effect of the intrinsic curvature on the Cauchy surface. Specifically, the large
intrinsic curvature allows for a relatively large number of quanta to be packed together in a spatial region with a radius on the order of the Schwarzschild radius
\footnote{In this case a spherical region with an area of
$\sim M^2$ can have a volume  $\sim M^4$, significantly larger than
what it would be if the intrinsic curvature were negligible.}.

Despite the fact that monsters saturate the Bekenstein-Hawking entropy bound
as initial data sets, they may not give rise to  valid gravitational collapse processes. In fact, it is far from clear if their full Cauchy development is
pathology free. This question is particularly important since a monster, by
virtue of being in the close vicinity of the future apparent horizon, is expected
to be largely within the future event horizon. Thus, one would need
to ascertain that the future event horizon is everywhere regular.

Moreover, monsters have a time symmetric Cauchy development. This
implies that they are endowed with two distinct curvature singularities,
a black hole and a white hole. 
White holes possess classical and quantum
instabilities that render them unphysical \citep{eardley,
  waldwhite,waldcpt}.
These issues suggest that monster initial sets are not physically
realistic or relevant
counterexamples to the conjectured relation
\eqref{genericrel}.

In this paper we investigate the global structure of
the full Cauchy development of monster-like initial conditions.
We shall focus on spacetimes with a dust dominated
collapsing core. We review the essential geometric
theory in Secs. \ref{monster},\ref{setup1}, and \ref{glue}.
In Sec. \ref{m2} we show how a monster-like initial condition
gives rise to a collapsing spacetime that saturates
the Bekenstein-Hawking entropy bound and has an everywhere
regular future event horizon.
Finally, in Sec. \ref{circ} we discuss a
class of examples obtained by modifying
the spacetime geometry in the past
in order to get rid of the white hole singularity.
We show that the modified spacetimes satisfy
the weak energy condition.

\section{the hydrodynamic approximation for dust fluids} \label{monster}

In this section we briefly review the basics of dust fluids and discuss
the conditions for the validity of the hydrodynamic approximation
in arbitrary spacetimes.

A pressureless dust fluid is characterized by a four-velocity vector field
$\vec{u}$ and a number density $n$, satisfying $u^{a} \nabla_{a} u^{b} = 0$ and
$ \nabla_a (n u^a ) =0$. The stress energy tensor for the dust fluid
is given by
\be \label{stress0}
T_{ab} = \rho \ u_a  u_b = m n \ u_a  u_b,
\ee
where $\rho = m n$ is the energy density measured in the comoving frame
of the fluid, and $m$ is the particle mass. Here we allow
$m$ to have dependence on spatial coordinates.

The entropy  $S$ of a dust fluid is roughly equal to the number
of its constituent particles. Therefore, one can associate a covariantly conserved entropy current $s^a = n u^a$ to the dust fluid
\footnote{the entropy current defined here is generally correct
upto some order unity prefactor.}, and
then compute the total entropy on any achronal spacetime slice $\Sigma$
by
\be \label{Entropy}
S = \int_{\Sigma} s^{a} \ d \Sigma_{a} = \int_{\Sigma} \ \frac{\rho}{m} u^{a} \ d\Sigma_a = \int_{\Sigma} \ n u^{a} \ d\Sigma_a.
\ee

The validity of the hydrodynamic approximation underlying the dust fluid model
hinges on the following conditions\footnote{These relations are not precise inequalities.
They are only defined upto factors of order unity. The numerical factors
are added for convenience when we later discuss saturating these relations.}:
\bea \label{validity}
&& (8 \pi m) \ n ^{-1/3} \gtrsim 1, \nonumber\\
&& (8 \pi m)^{-1} \gtrsim 1, \nonumber\\
&& \Big(\frac{16 \pi}{3} \Big)^{1/3}\frac{m}{|\bold{D} m|} \gtrsim n^{-1/3}, \nonumber\\
&& \mathcal{L} \gg n^{-1/3},
\eea
where $\bold{D}$ denotes differentiation with respect to
the spatial coordinates, and $\mathcal{L}$ is the radius of curvature
defined as $\mathcal{L}^{-4} \equiv R_{abcd} R^{abcd}$ for the spacetime
Riemann curvature tensor $R_{abcd}$. To have a valid model of gravitational
collapse, one needs to ensure that these relations are valid everywhere
to the past of the event horizon of the ensuing black hole.

The above conditions can be understood as follows. The first condition
is the non-degeneracy condition, asserting that the degeneracy pressure
can be ignored as long as the dust particles remain non-overlapping. The
violation of this condition undermines the pressureless dust description of the fluid
\footnote{As the intra-particle spacing approaches the Compton wavelength
of particles, the quantum fluctuations in the stress energy tensor
of the dust fluid become comparable to the classical stress energy tensor.}.
 The second
condition asserts that the Compton wavelength of each particle must be
above the Planck length \footnote{The Planck length is defined as $\sqrt{\hbar G/c^3}$, which is equal to unity in our choice of units.} for quantum gravity effects to be ignored.
The third and fourth conditions require the radius of curvature
and the length scales over which the particle mass varies to be
larger than the intra-particle spacing.  This last
condition is equivalent to saying that most of the contribution
to the total entropy comes from the modes with wavelengths small compared
to the spacetime radius of curvature.

\section{\label{setup0} dust objects in spherical symmetry}
\subsection{\label{setup1} Lema\^itre-Tolman-Bondi dust models}

Consider a spherically symmetric compact object that is composed of non-interacting
dust particles. For an arbitrary density profile \footnote{For homogeneous density profiles, the metric
is the Friedmann-Robertson-Walker (FRW) metric
$$ds^2 = -dt^2 + a(t)^2 \bigg[ \frac{dr^2}{1-\kappa r^2} + r^2 d \Omega^2 \bigg]$$
where $\kappa$ is a constant taking $0,\pm1$ values.
The areal radius $R(t,r)$ in the metric \eqref{LTB} then reduces to $r a(t)$ in the homogeneous case.}, the interior geometry of the object is described by the Lema\^itre-Tolman-Bondi (LTB) metric
\citep{bondi,celerier,vanderveld}
\be \label{LTB}
ds^2 = - dt^2 + \frac{R'(t,r)^2 dr^2}{1-r^2 k(r)} + R(t,r)^2 d \Omega^2,
\ee
where prime denotes differentiation with respect
to the coordinate $r$, $R(t,r)$ is the areal radius, $k(r)$ determines
the intrinsic curvature on constant $t$ slices
\footnote{The intrinsic scalar curvature of
the constant $t$ slices is
${}^{(3)} R = 6 k(r) +2 r k'(r)$.},
and $d\Omega^2 = d \theta ^2 + \sin{\theta}^2 d \phi ^2$.
Since the metric \eqref{LTB} is describing the interior of a compact
region, the coordinate radius $r$ is only defined out to some $r_{\text{out}}$.
Furthermore, $R'(t,r) >0$ and $k(r)<1/r^2$ as the spherical shells are not allowed to cross one another.

Given the metric \eqref{LTB}, the stress energy tensor
for the pressureless dust fluid is given by \eqref{stress0}
with
\be \label{stress}
\vec{u} = \vec{\partial}_{t}, \ \ \rho(t,r) =
\frac{ \bar{\rho} r^2}{ R(t,r)^2 \  R'(t,r)},
\ee
for some positive constant
$\bar{\rho}$. The metric function $R(t,r)$ can be solved for analytically from
the $G_{rr} = 8 \pi T_{rr} =  0$ component of the Einstein's equations. For the reason mentioned in footnote 5, we will restrict attention
to the $k(r) >0$ case. In this case we have
\be \label{ltbeq}
G_{rr} = 0 \Rightarrow \dot{R}(t,r)^2 + 2 R(t,r) \ddot{R}(t,r) + r^2 k(r) = 0,
\ee
where dot denotes differentiation with respect to $t$.
The closed form solution to Eq. \eqref{ltbeq} is given
by the parametric equations
\be \label{sol}
R(t,r) = \frac{4 \pi \bar{\rho} r}{3 k(r)} (1-\cos{u}), \ \ \ t-t_{0} (r) = \frac{4 \pi \bar{\rho} }{3 k(r)^{3/2}}(u-\sin{u}),
\ee
where $0 \leq u \leq 2\pi$. Evidently, there are two curvature
singularities in this spacetime. The function $t_0(r)$ is the "bang function"
specifying the coordinate time at which a spherical shell at a coordinate
radius $r$ departs from one of the curvature singularities, only
to arrive at the other one at the coordinate time $t = t_0(r) + 
4 \pi ^2 \bar{\rho} / [3 k(r)^{3/2}]$.
Note that the function $t_0 (r)$ is chosen such that
the inner spherical shells arrive at the curvature singularities
prior to the outer spherical shells.

\subsection{Dust objects with vacuum exterior}\label{glue}

In this section we construct models of collapsing dust objects
by gluing LTB interior solutions to Schwarzschild exterior solutions.

As we mentioned in Sec. \ref{setup1}, the LTB metric describes
the interior geometry of a compact spherical region.
The boundary of the interior region is a timelike three dimensional
hypersurface that consists of points with the LTB coordinates $\{t,r_{\text{out}}\}$.

The exterior geometry is given by
the Schwarzschild metric
\be \label{schwarz}
ds^2 = - \omega(\bar{r}) d\bar{t} ^2 +
 \omega(\bar{r}) ^{-1} d\bar{r} ^2 +
 \bar{r}^2 d\Omega^2, \ \ \omega(\bar{r}) = 1-\frac{2M}{\bar{r}},
\ee
where $M$ is the gravitational mass associated with the collapsing object.
The boundary of the compact inner region in the Schwarzschild coordinates
is given by $\big\{ t,R(t,r_{\text{out}}) \big\}$, where $t$ is the
proper time of the freely falling dust particles on the boundary
of the inner region. Note that $t$ coincides with the LTB time coordinate.

To smoothly join inner LTB solutions to outer Schwarzschild
solutions one must impose the Israel junction conditions on
the boundary. The details of this process is essentially identical to the
Oppenheimer-Snyder collapse and it can be found in many references [e.g.
see \citep{MTW}]. It follows from imposing the Israel junction
conditions on the boundary region that
\bea \label{adm}
&& M = \frac{R(t,r_{\text{out}})}{2} \bigg[ \dot{R}(t,r_{\text{out}}) ^2 + r_{\text{out}}^2 k(r_{\text{out}}) \bigg]\nonumber\\
&& = \frac{4 \pi \bar{\rho} r^3_{\text{out}}}{3},
\eea
where the second equality above follows from Eqs.\eqref{ltbeq} and \eqref{sol}.  Actually, a more general statement follows from Eq.\eqref{ltbeq};
that the Misner-Sharp mass enclosed by each spherical shell at
the LTB coordinate radius $r$ is 
\be \label{misner}
M(r) = \frac{R(t,r)}{2} \bigg[ \dot{R}(t,r) ^2 + r^2 k(r) \bigg] = \frac{4 \pi \bar{\rho}  r^3}{3}.
\ee
The future (and past) apparent horizon is then
defined as the surface for which
\footnote{This can also be derived by finding the necessary condition for which the congruences of the outgoing future and past directed
null geodesics that are orthogonal to a 2-sphere at the coordinate radius $r$
have zero expansions.}
\footnote{It is proven in \citep{ellis} that for strongly predictable spacetimes satisfying the weak or strong energy conditions the future apparent horizon is inside
the future event horizon. The LTB-Schwarzschild spacetimes for which
the future event horizon is regular everywhere are
strongly predictable and satisfy both the weak and strong energy conditions.}
\be \label{apparent}
1- \frac{2 M(r)}{R(t,r)} = 1- r^2 k(r) - \dot{R}(t,r)^2 = 0
\ee
for all $r$. All spherical shells are said to be marginally trapped on the apparent horizon. Once the apparent horizon is crossed, i.e.
\be \label{apphori}
1- \frac{2 M(r)}{R(t,r)} = 1- r^2 k(r) - \dot{R}(t,r)^2 < 0
\ee
for some spherical shell at $r$, then the shell is said to have become
trapped.

\subsection{Dust objects with order $M^2$ entropy} \label{m2}

In this section we show that one can construct collapsing dust objects
using the LTB geometries with $k(r)>0$ that come close to saturating the
Bekenstein-Hawking entropy bound of $S_{\text{BH}} = 4 \pi M^2$
for non-black hole objects
\footnote{This cannot be done for homogeneous dust
objects in the Oppenheimer-Snyder collapse. The entropy \eqref{Entropy}
in these models is roughly $M/m$. However, for the fluid
to obey the first and fourth of the conditions given in \eqref{validity} we
must have $m^{-1} \lesssim n^{-1/3} \ll M$. Similar conclusions hold for the LTB
geometries with $k(r)=0$ or $k(r)<0$.}.

Using Eqs. \eqref{Entropy} \footnote{Note that the LTB metric
\eqref{LTB} together with Eq. \eqref{sol} guarantee $\nabla_{a}(\rho u^a /m) = 0$ for $m$ being an arbitrary function of $r$.}
\eqref{LTB} (for $k(r)>0$) and \eqref{stress} we
find the entropy of a collapsing dust object on an achronal
but otherwise
arbitrary slice to be
\bea \label{entropy2}
&&S = 4 \pi \int_0^{r_{\text{out}}}\ dr \frac{\bar{\rho} r^2}{m(r)\sqrt{1-r^2 k(r)}}.
\eea
Clearly, not all choices of the metric and particle mass functions
result in valid models that can saturate the Bekenstein-Hawking entropy bound.
In addition to requiring the hydrodynamic relations \eqref{validity} to remain valid
everywhere to the past of the future event horizon \footnote{For some choices of the LTB metric functions, it is possible for the future event horizon to encounter the past curvature singularity
at some $r>0$ [see App. \ref{nocauchy}]. It is then implied that the hydrodynamics conditions
\eqref{validity} are violated somewhere on the future event horizon. Additionally, it follows that there are no spacetime slices that fall entirely outside of the future event horizon. Thus, the integral \eqref{entropy2}
does not account for the entropy of a non-black hole object.}, 
we must also require most of the future apparent horizon 
to consist of  2-spheres for which $|\dot{R}(t,r)|\ll1$.
This is a revision of a similar requirement
that was originally set forth in \citep{sorkin}. It can be seen
from Eq. \eqref{apparent} that the future (or past) apparent horizon
cannot be entirely foliated by 2-spheres with $\dot{R}=0$. 
In fact that would require
$k(r) = 1/r^2$ for all $r$, for which the metric \eqref{LTB} would not be defined. Nonetheless, one might wonder if there are any constraints on
how small $|\dot{R}|$ is for the 2-spheres that foliate the apparent horizon. Without any constraints, the integral \eqref{entropy2} can be made arbitrarily large for choices of $k(r)$ that are arbitrarily close to $1/r^2$.

A physically realistic choice of constraint on the smallness of $|\dot{R}|$
along the apparent horizon can be
explained as follows. Consider the 3-dimensional achronal surface
$\Sigma$ parametrized by $t_{\Sigma}(r)$ 
that is foliated by the 2-spheres with $\dot{R}=0$. 
Now consider a constant $t$ surface that intersects 
$\Sigma$ at $\{t_{\Sigma}(\check{r}),\check{r}\}$.
Define a coordinate radius $\tilde{r}$ on this constant
$t$ surface using the following relation,
\be 
R[t_{\Sigma}(\check{r}), \tilde{r}] = 2 M(\check{r}).
\ee
The proper distance between the shell at $\check{r}$ and 
the shell at  $\tilde{r}$ on this constant 
$t$ surface is given by 
\bea 
 \int_{\tilde{r}}^{\check{r}} \ \sqrt{g_{rr}} dr.
\eea
This distance corresponds to the spatial distance that the shell
at $\{t_{\Sigma}(\check{r}),\check{r}\}$ would have to travel 
in order to arrive at the apparent horizon. We require this
distance to be larger than the particle spacing on the shell at $\check{r}$, i.e. 
\bea \label{shell-apphori}
 \int_{\tilde{r}}^{\check{r}} \ \sqrt{g_{rr}} dr \geq 4 n^{-1/3} \big[t_{\Sigma}(\check{r}),\check{r}\big],
\eea
\footnote{The factor $4$ is added so that the equality
case for the relation \eqref{shell-apphori} results in the entropy
\eqref{entropy2} being bounded by $S_{\text{BH}}$}.
According to the this constraint, $\Sigma$ cannot be arbitrarily 
close to the apparent horizon since that would require 
the number density $n$ to diverge on $\Sigma$,
which takes the dust fluid outside of its regime of validity.
We shall impose this constraint in the example that we discuss below.

We now construct an example as follows. We set $8 \pi \bar{\rho} =1$
and assume that $0 \leq r \leq 99/10$
for the interior region. It follows from Eq. \eqref{adm} that
the collapsing object has a gravitational mass $M= 161.7$ and a
Bekenstein-Hawking entropy
$S_{\text{BH}} = 3.3 \times 10^{5}$ for the ensuing black hole in
the Planck units.
We then require the spacetime to admit a null Cauchy surface $\Sigma$
which is foliated by the spheres with $\dot{R} = 0$ in the interior region,
i.e. $u = \pi$ everywhere on the
interior part of $\Sigma$. Evaluating
Eq. \eqref{sol} at $u= \pi$ we find
\be \label{solpi}
R\big[t(r),r\big] = \frac{ r}{3 k(r)}, \hspace{1cm} t(r) = t_0(r) + \frac{ \pi}{6 k(r)^{3/2}},
\ee
the latter giving the defining equation for the surface $\Sigma$ in the
interior region.
We can then use the fact that $\Sigma$ is a null surface together
with Eqs. \eqref{LTB}, \eqref{sol} and \eqref{solpi} to find a differential
equation for the bang function $t_0 (r)$:
\bea \label{bang}
&&\frac{d t}{dr} = \frac{R'(t,r)}{\sqrt{1-r^2 k(r)}} \bigg |_{\Sigma}\nonumber\\
&&\Rightarrow t_0 '(r) = \frac{1}{3 k(r)} \Bigg[ \frac{1- \frac{r k'(r)}{k(r)}}{\sqrt{1-r^2 k(r)}} + \frac{3 \pi}{4} \frac{k'(r)}{k(r)^{3/2}} \Bigg].
\eea
Thus, the knowledge of $k(r)$ is sufficient to fully specify the interior geometry. Once we have the interior metric, we can find the
location of the future event horizon by first solving Eq. \eqref{sol}
for $r_{\text{out}}$ to find the time $t_{*}$
when the outermost spherical shell becomes trapped, and then
solve the following equation for the null geodesic in the interior region with $\{t_{*},r_{\text{out}} \}$ as its
end point:
\bea \label{event}
&&\frac{d t}{dr} = \frac{R'(t,r)}{\sqrt{1-r^2 k(r)}} \nonumber\\
&& = \frac{\partial_{r} \Bigg[\frac{r}{6 k(r)} \Big(1- \cos\big[\mathcal{F}\big \{6 k(r)^{3/2} (t-t_0 (r) \big \}\big]\Big) \Bigg]}{\sqrt{1-r^2 k(r)}},
\eea
where we defined $\mathcal{F}(u-\sin u) \equiv u$ and used Eq. \eqref{sol}.
We could then confirm that the surface $\Sigma$ is indeed located
outside of the future event horizon.

\begin{figure}[t]
\centering
\includegraphics[width=0.5\textwidth]{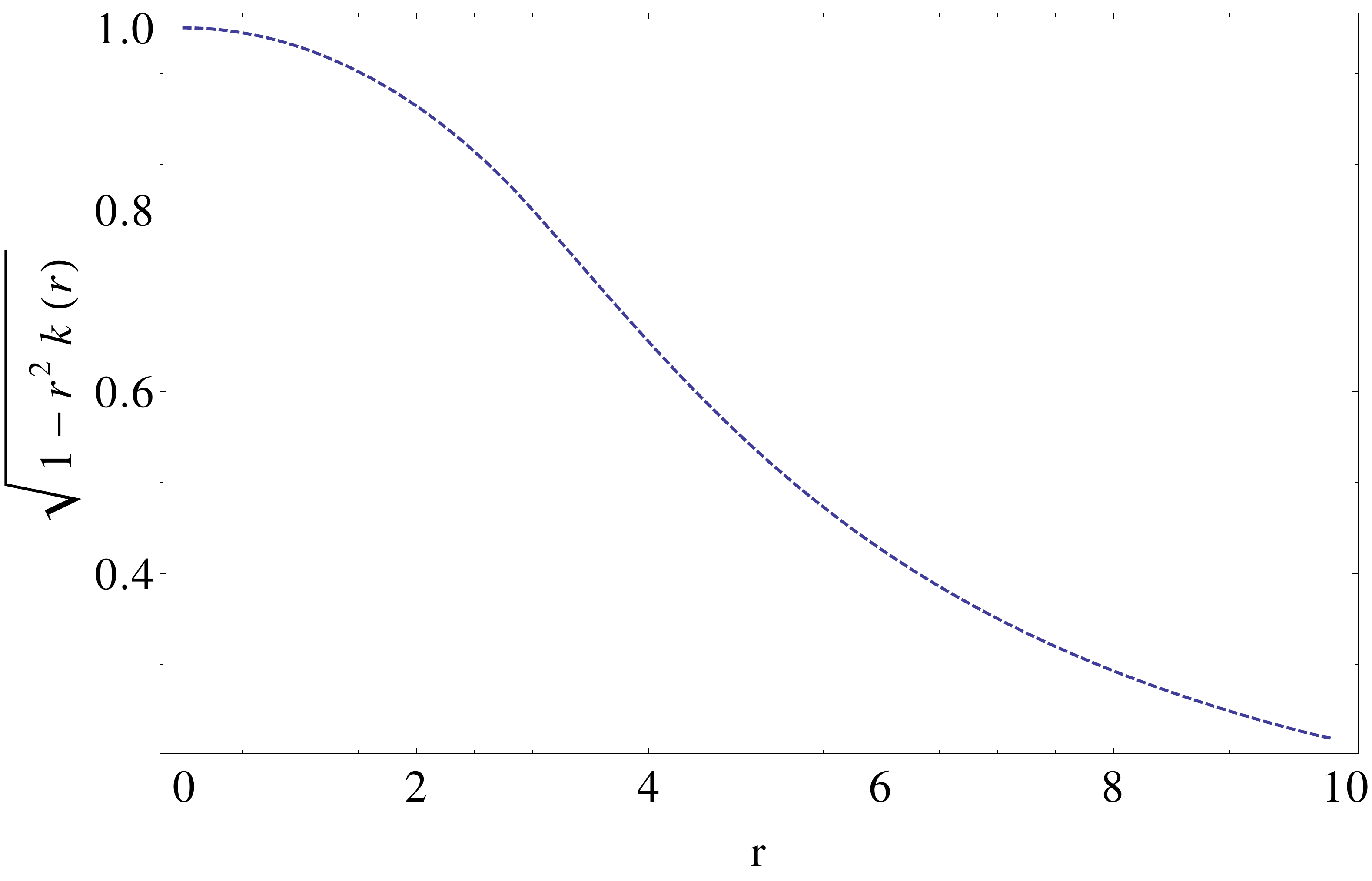}
\caption{This is the plot of $\sqrt{1-r^2 k(r)}$ versus $r$.
As expected, $1-r^2 k(r) \ll 1$ for $7 \leq r \leq 99/10$.}
 \label{intcurv}
\end{figure}

We proceed by finding a differential equation for $k(r)$ by saturating both the non-degeneracy condition given in \eqref{validity} and the shell physical distance condition given in \eqref{shell-apphori} for
some spherical shells on $\Sigma$. On a given constant $t$ surface
intersecting $\Sigma$ at $\{t_{\Sigma}(\check{r}),\check{r}\}$, we have
\bea \label{saturated}
&&\int_{\tilde{r}}^{\check{r}} \ \sqrt{g_{rr}} dr = \int_{\tilde{r}}^{\check{r}}  \ \frac{R'\big[t(r),r \big] dr}{\sqrt{1-r^2 k(r)}} dr \nonumber\\
&&  \approx \frac{1}{\sqrt{1-\check{r}^2 k(\check{r})}}\int_{\tilde{r}}^{\check{r}} R'\big[t(r),r \big] dr \nonumber\\
&& = \frac{R\big[t(\check{r}),\check{r} \big]- 2 M[\check{r}]}{\sqrt{1-\check{r}^2 k(\check{r})}} = R\big[t(\check{r}),\check{r}\big]\sqrt{1-\check{r}^2 k(\check{r})}   \nonumber\\
&& = 4 \times \big(8 \pi\rho[ t(\check{r}),\check{r}]\big)^{-1/4} ,
\eea
where we approximated the integral by assuming that
the shell at $\{t(\check{r}),\check{r}\}$ is nearly trapped, i.e. $R\big[t(\check{r}),\check{r} \big]- 2 M(\check{r})
= 1- \check{r}^2 k(\check{r}) \ll 1$.
We construct $k(r)$  by
solving Eq. \eqref{saturated} for $7 \leq r \leq 99/10$ and then continuously
joining that solution to a slowly varying function of $r$
for $0 \leq r \leq 7$
\footnote{The resulting $k(r)$ need not necessarily be $C^{\infty}$ but
it should be at least $C^2$.}. We plot the resulting $\sqrt{1-r^2 k(r)}$ in
Fig.\ref{intcurv}. Note that $1-r^2 k(r) \ll 1 $ for $7 \leq r \leq 99/10$
is consistent with the assumption $R\big[t(r),r \big]- 2 M(r) \ll 1$ 
for these shells used in Eq. \eqref{saturated}.

\begin{figure}[t]
\centering
\includegraphics[width=0.5\textwidth]{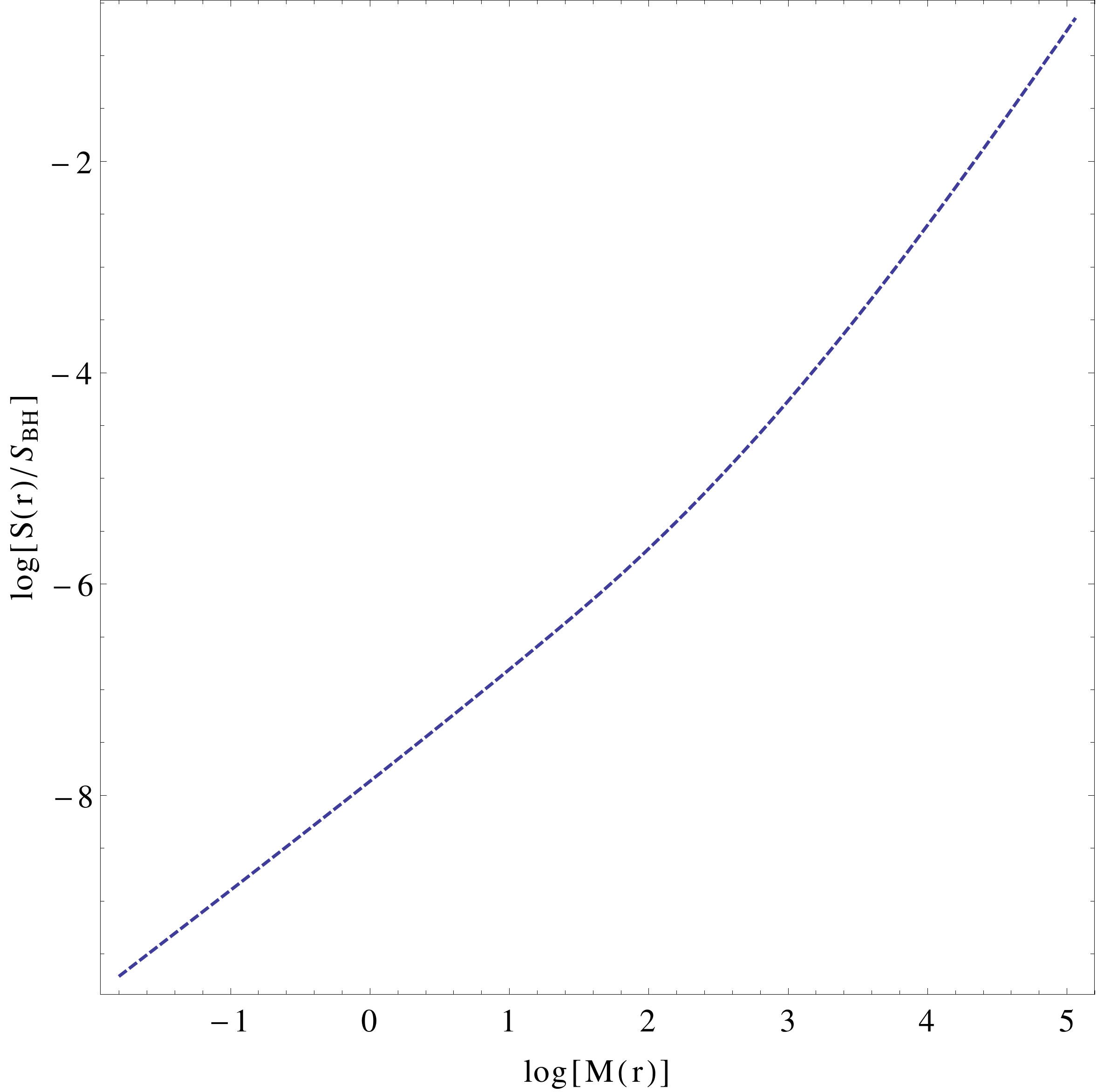}
\caption{This is the plot of $\log \big[S(r)/S_{\text{BH}}\big]$ versus
$\log \big[M(r) \big]$ for the example we have constructed here. Notice
that the contribution of the spherical shells at $r \geq 7$ brings
the entropy given in Eq. \eqref{entropy2} close to
the Bekenstein-Hawking entropy $S_{\text{BH}} = 3.3 \times 10^{5}$.}
\label{entropyratio}
\end{figure}

\begin{figure}
\centering
\includegraphics[width=0.5\textwidth]{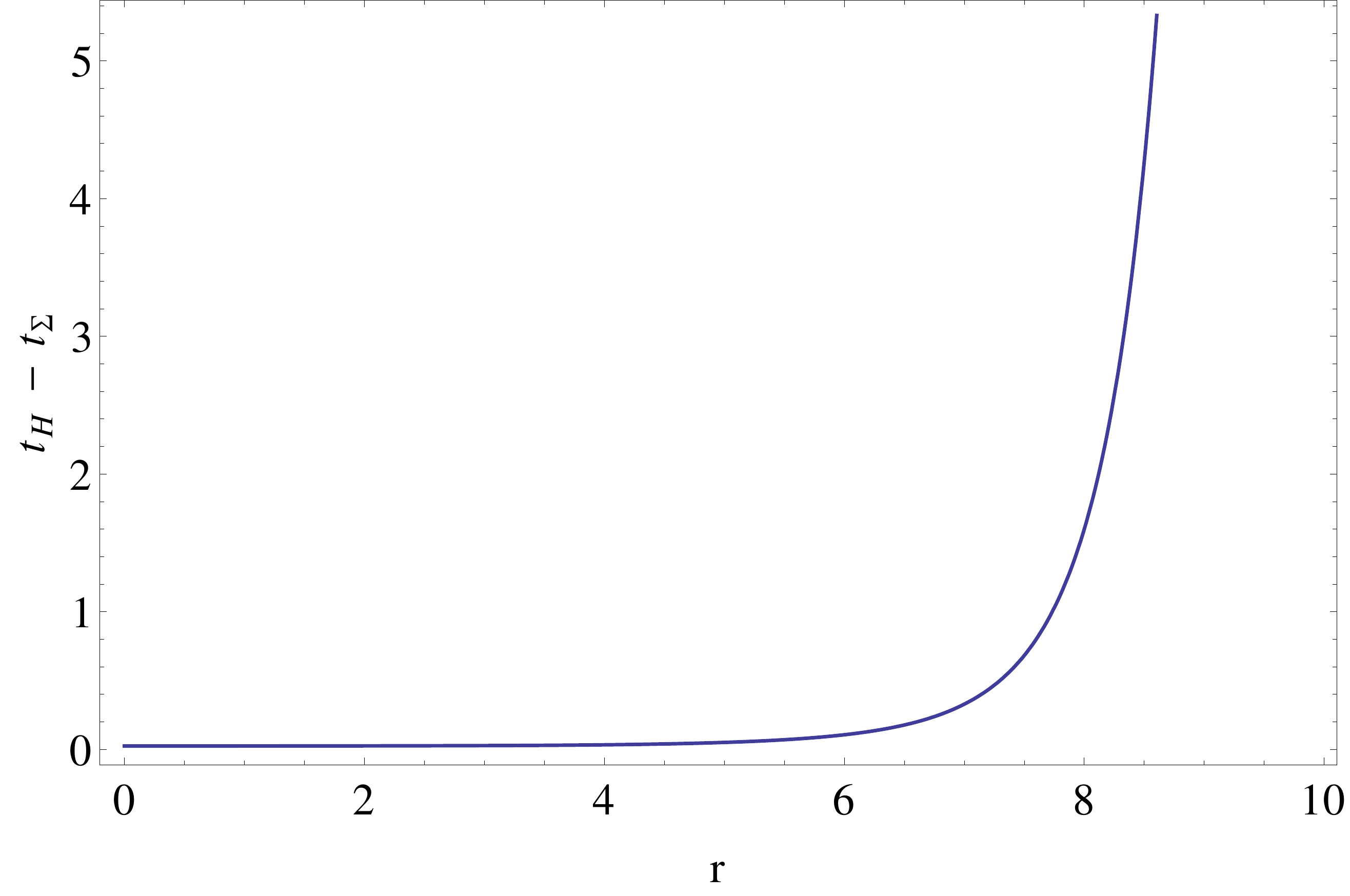}
\caption{This is the plot of $t_{\text{H}} - t _{\Sigma} $
versus radius $r$. The future event horizon is indeed to the future of
the slice $\Sigma$}
\label{sig-eve}
\end{figure}

\begin{figure}
\centering
\includegraphics[width=0.5\textwidth]{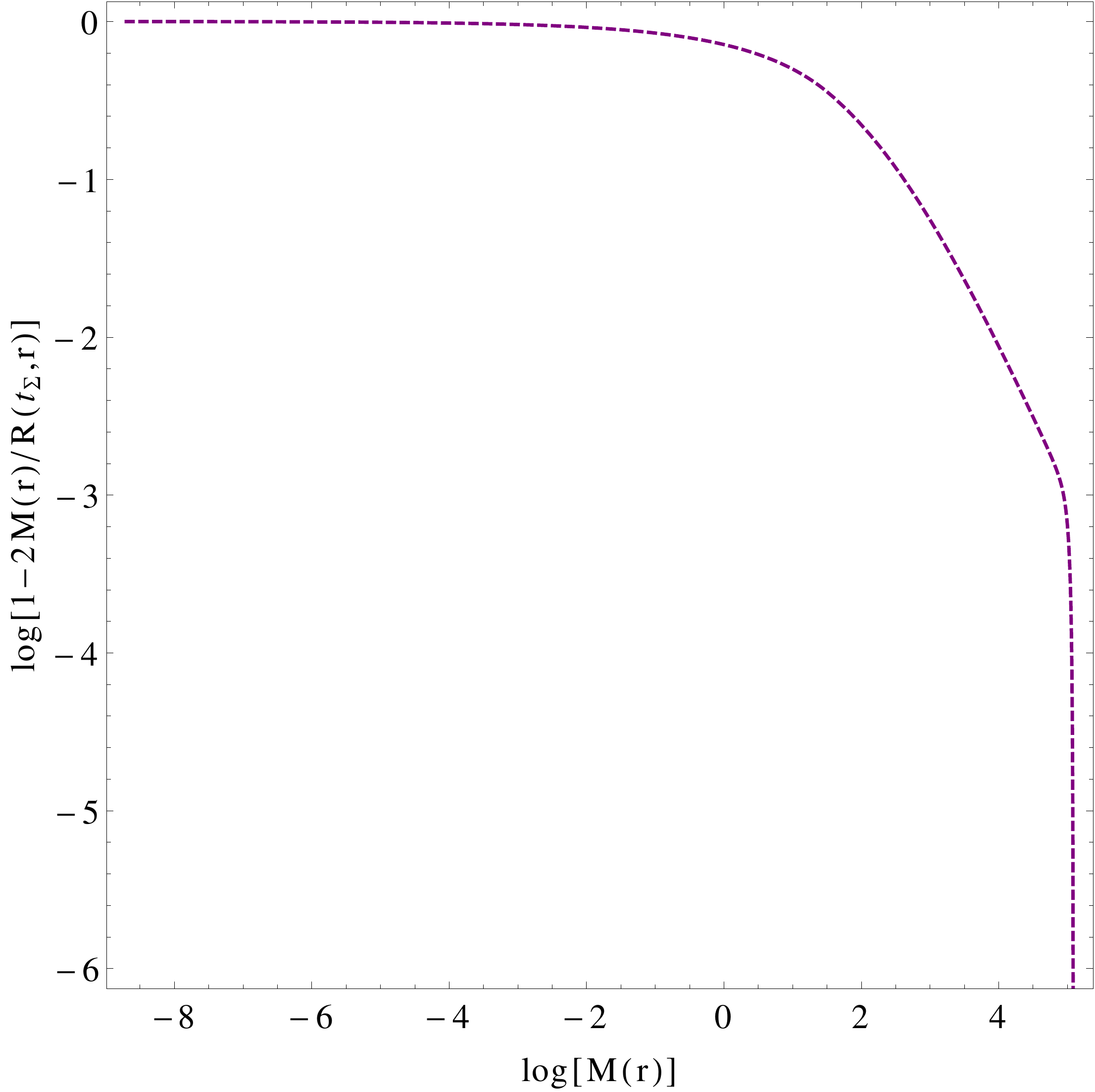}
\caption{Here we plot $\log [1-2M(r)/R(t_{\text{H}},r)]$ verses $\log [M(r)]$
along the future event horizon. Notice that none of the spherical
shells are trapped, except for the outermost one for which
$\log [1-2M(r_{\text{out}})/R(t_{\text{H}},r_{\text{out}})]$ diverges.}
\label{event-trapped}
\end{figure}

\begin{figure}
\centering
\includegraphics[width=0.5\textwidth]{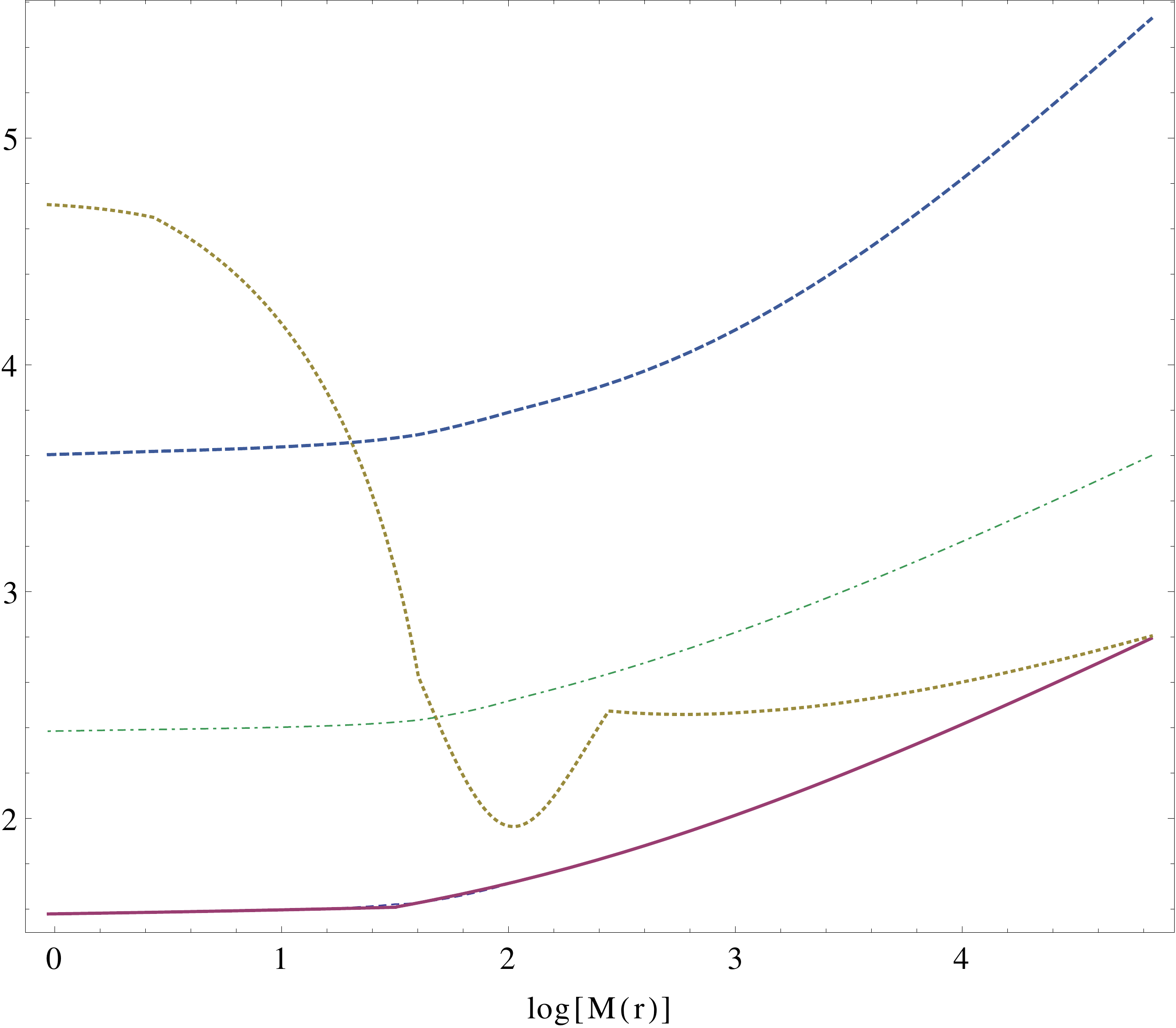}
\caption{This is the plot of $\log \big[n^{-1/3} \big]$ (the solid blue line), $\log \big[(8 \pi m)^{-1} \big]$ (the solid violet line which is collinear
with the solid blue line),
$\log \big[(16 \pi/3)^{1/3} m/|\bold{D} m| \big]$ (the dotted brown line),
$\log \big[\rho^{-1/4} \big]$ (the green dotted-dashed line),
and $\log \big[\mathcal{L} \big]$ (the blue dashed line) versus
$\log \big[M(r) \big]$ along the future event horizon.}
\label{hydro}
\end{figure}

With $k(r)$ in hand, we can now compute the integral \eqref{entropy2}
to find the total entropy
\be
S= 1.8 \times 10^{5} \approx 0.55 \ S _{\text{BH}},
\ee
which turns out to be quite close to saturating the Bekenstein-Hawking
entropy bound. See Fig. \ref{entropyratio}

Upon solving Eqs. \eqref{bang} and \eqref{event},
we find the location $t_{\Sigma}$ of the slice $\Sigma$ as well as
the location $t_{\text{H}}$ of the future event horizon.
We plot $t_{\text{H}} - t _{\Sigma} $ in Fig. \ref{sig-eve}.
Note that the surface $\Sigma$ is completely outside of the future event horizon.
Additionally, as expected, the future apparent horizon is inside the future event
horizon. See Fig. \ref{event-trapped}.

Finally, in Fig. \ref{hydro} we plot the four relevant fluid
micro-length scales together with the radius of curvature,
$\{n^{-1/3},(8 \pi m)^{-1},(16 \pi/3)^{1/3} m/|\bold{D} m|, \rho^{-1/4}, \mathcal{L} \}$,
defined in Sec. \ref{monster} along the future event horizon. Notice that
all micro-length scales are larger than the particle spacing $n^{-1/3}$ everywhere along the future event horizon. Notice that the third condition
given in \eqref{validity} is close to becoming saturated.
This indicates that the length scales over which the particle mass function $m(r)$ varies is becoming comparable
to the particle spacing $n^{-1/3}$ along the future event horizon.
Also, one should note that all five length scales are above the Planck
length.

\section{attempts to find past geometries with no initial singularity}\label{circ}

In Sec. \eqref{m2} we constructed a collapsing dust object
that saturates the Bekenstein-Hawking entropy bound.
However, as we mentioned in the introduction, the global time symmetry
of the resulting spacetime renders it pathological due to the presence
of a white hole.  We find it worthwhile to attempt to construct
spacetimes that have the same large entropy but not the white hole
singularity in the past.

In the remainder of this paper, we report on a formalism
that we have explored in order to circumvent the past white
hole singularity. We find it convenient to initially
lay out the formalism for the closed FRW geometries, and then apply it to the LTB geometries with $k(r)>0$. As we show below, the numerical
investigations appear to be indicative of the success of this
formalism in both avoiding the white hole singularity and
keeping the interior modified spacetime's stress energy tensor from violating
the dominant energy condition \footnote{See \citep{waldgr} for a definition
of the energy conditions.}. However, we show in Sec. \ref{modltb}
that a boundary stress energy tensor is needed to match the 
interior metric to the exterior Schwarzschild geometry. Given that
this boundary stress energy tensor only satisfies the weak energy
condition, the overall spacetime only satisfies the weak energy
condition as well.


\subsection{The modified closed Friedman-Robertson-Walker geometry} \label{formalism}

In this section we show that one can modify the closed FRW geometry
in the past in order to avoid the white hole singularity. By construction,
the stress energy tensor of the modified spacetime satisfies the dominant
energy conditions.

The metric for the modified spacetime is
\be \label{metric2}
ds^2 = a(\eta)^2 \bigg[-d\eta^2 + d \chi ^2  +\bigg(\frac{\sin{ \sqrt{\kappa(\eta)} \chi}}{\sqrt{\kappa(\eta)}}\bigg)^2 d\Omega^2 \bigg].
\ee
where $\eta \equiv \int dt/a(t)$ is the conformal time and $\chi \equiv \int dr/\sqrt{1-\kappa r^2}$. Here, unlike the closed FRW geometry, the spatial curvature $\kappa$ is a function of the conformal time $\eta$ \footnote{See footnote 10.}.
We require $\kappa(\eta) > 0$ at all times and $\kappa(\eta)= 1$
for $\eta \geq \eta_{\Phi}$, where $\eta_{\Phi}$ is a conformal
time beyond which the spacetime geometry reduces to the closed FRW geometry.
For times $-\infty<\eta\leq \eta_{\Phi}$, we define the metric functions $a(\eta)$
and $\kappa(\eta)$ by the following set of equations
\bea \label{flow}
&& 2 \mathcal{H}'+ \mathcal{H}^2+ \kappa \bigg(1+ \frac{\mathcal{L}}{\mathcal{H}} \bigg)=0, \nonumber\\
&& \mathcal{L}'- \frac{1}{2} \mathcal{L}^2+\mathcal{H}\mathcal{L} = 2 g \mathcal{H}',
\eea
with the initial conditions
\be
\kappa(\eta_{\Phi})=1, \hspace{1cm}\kappa '(\eta_{\Phi})=0,\hspace{1cm}0 < \mathcal{H}(\eta_{\Phi}) \ll 1 \footnote{Note that we are evolving the metric
functions from $\eta_{\Phi}$ to $-\infty$. Therefore, the sign of $\mathcal{H}(\eta_{\Phi})$ is positive as the spacetime is expanding in this direction.}.
\ee
The functions appearing in Eq. \eqref{flow}
are $\mathcal{L}\equiv \kappa'/\kappa$, $\mathcal{H} \equiv a'/a$, and
$g(\eta) \equiv 1- \exp{[-(\eta_{\bar{\Phi}}-\eta)]}$
for $\eta \leq \eta_{\bar{\Phi}}$ and zero otherwise.
One might consider
other choices of $g(\eta)$ depending on the smoothness
properties that one wishes to demand of the stress energy tensor. For this particular
choice of $g(\eta)$, the stress energy tensor will be continuous at $\eta = \eta_{\bar{\Phi}}$
and smooth everywhere else.
Geometrically, we start from a point near the moment of time symmetry
for the closed FRW geometry and evolve the spacetime backwards in time
while diluting the spatial curvature on constant $\eta$ slices.
The objective is to push the moment of time symmetry off to $\eta \rightarrow - \infty$, thereby avoiding the white hole singularity.

Before discussing the properties of the modified FRW spacetime,
we find it helpful to comment on the motivation behind the set of
equations given in \eqref{flow}. The first equation can be regarded
as the conservation of the homogenized Misner-Sharp mass. 
Indeed if we define $R(t,r) = r a(t)$ and require $k$ to have
time dependence in Eq. \eqref{misner}, we arrive at the  first equation in  \eqref{flow} by setting $M_{,t}=0$. The second equation in \eqref{flow}
is perhaps less motivated, mainly designed to give a sufficiently
slow evolution for $\kappa$ in order to keep the modified spacetime
from violating the dominant energy condition.

We now provide more details on the solutions to the dynamical
equations given in \eqref{flow}. 
To see that the moment of time symmetry is avoided at any finite
time, note that initially $\mathcal{H}'(\eta_{\Phi}) <0$ and $\mathcal{H}(\eta_{\Phi})>0$ as we evolve the spacetime backwards in time.
If $\mathcal{L}<0$ and $\kappa>0$ at all times $\eta < \eta_{\Phi}$, then
$\mathcal{H}$ cannot go to zero since that would imply
$\mathcal{H}'\rightarrow +\infty$, which is a contradiction.

To see how $\mathcal{L}$ evolves in time, note that the second equation
in \eqref{flow} together with the initial conditions
$\mathcal{L}(\eta_{\Phi})=\mathcal{L}'(\eta_{\Phi}) =0$ implies that
$\mathcal{L}''(\eta_{\Phi}) < 0$. Thus, $\mathcal{L}<0$ at times $\eta<\eta_{\Phi}$ and sufficiently close to $\eta_{\Phi}$.
Furthermore, we expect to have $\mathcal{L} > -\mathcal{H}$ as long as $\kappa > 0$.
Indeed if $\mathcal{L} = - \mathcal{H}$ at some finite time $\eta_{*} < \eta_{\Phi}$, Eq. \eqref{flow} implies
$\mathcal{L}'(\eta_{*}) > |\mathcal{H}'(\eta_{*})| = \mathcal{H}(\eta_{*})^2/2 $. Therefore,
while $\mathcal{L}$ initially decreases below $0$ we must
have $\mathcal{L} > - \mathcal{H}$. The only way that $\mathcal{L}$
can become equal to $-\mathcal{H}$ at some time $\eta_{*}$ is to
have $0 \leq \mathcal{L}' (\eta_{*}) \leq |\mathcal{H}'(\eta_{*})|$
which is inconsistent. Likewise, $\mathcal{L}(\eta_{*})=0$ is forbidden
since it requires $\mathcal{L}'\geq0$ in the vicinity of $\eta_{*}$,
whereas Eq. \eqref{flow} implies $\mathcal{L}' = 2 g \mathcal{H}'<0$ at $\eta_{*}$.

 In fact, given our choice of initial conditions,
it is not difficult to see that the dynamical system under study
has an attractor solution $\mathcal{L} \rightarrow -\mathcal{H}$.
Informally, we can see that this solution is a true attractor by writing
$\mathcal{L}(\eta) = \mathcal{L}_{0} (\eta) + \delta (\eta) $, where $\mathcal{L}_0 = -\mathcal{H}$ and initially $|\delta| \ll |\mathcal{L}_0|$. We
can then substitute this expression for $\mathcal{L}$  into the second equation in \eqref{flow} and assume $|\eta-\eta_{\Phi}| \gg 1$ to find the following differential equation for $\delta$
\be \label{delta}
\delta_{,\eta} - \frac{1}{2} \delta^2 - 2 \mathcal{L}_0 \delta = 0.
\ee
Since for the attractor solution we have $\mathcal{H}  \approx -\mathcal{L} \approx 2/|\eta|$, Eq. \eqref{delta} gives $\delta \propto 1/\eta^4$, which falls off faster than $\mathcal{L}_0$ for large $|\eta|$ as expected. Thus, we have $a \propto \eta^2$ and $\kappa \propto 1/ \eta^2$ as $\eta \rightarrow - \infty$. As a result, one can check that the components of the Riemann tensor for the constant $\eta$ slices go to zero as $\eta \rightarrow -\infty$. This is equivalent
to saying that the constant $\eta$ slices become nearly intrinsically flat at
large $|\eta|$.

\begin{figure}[h]
\centering
\includegraphics[width=0.5 \textwidth]{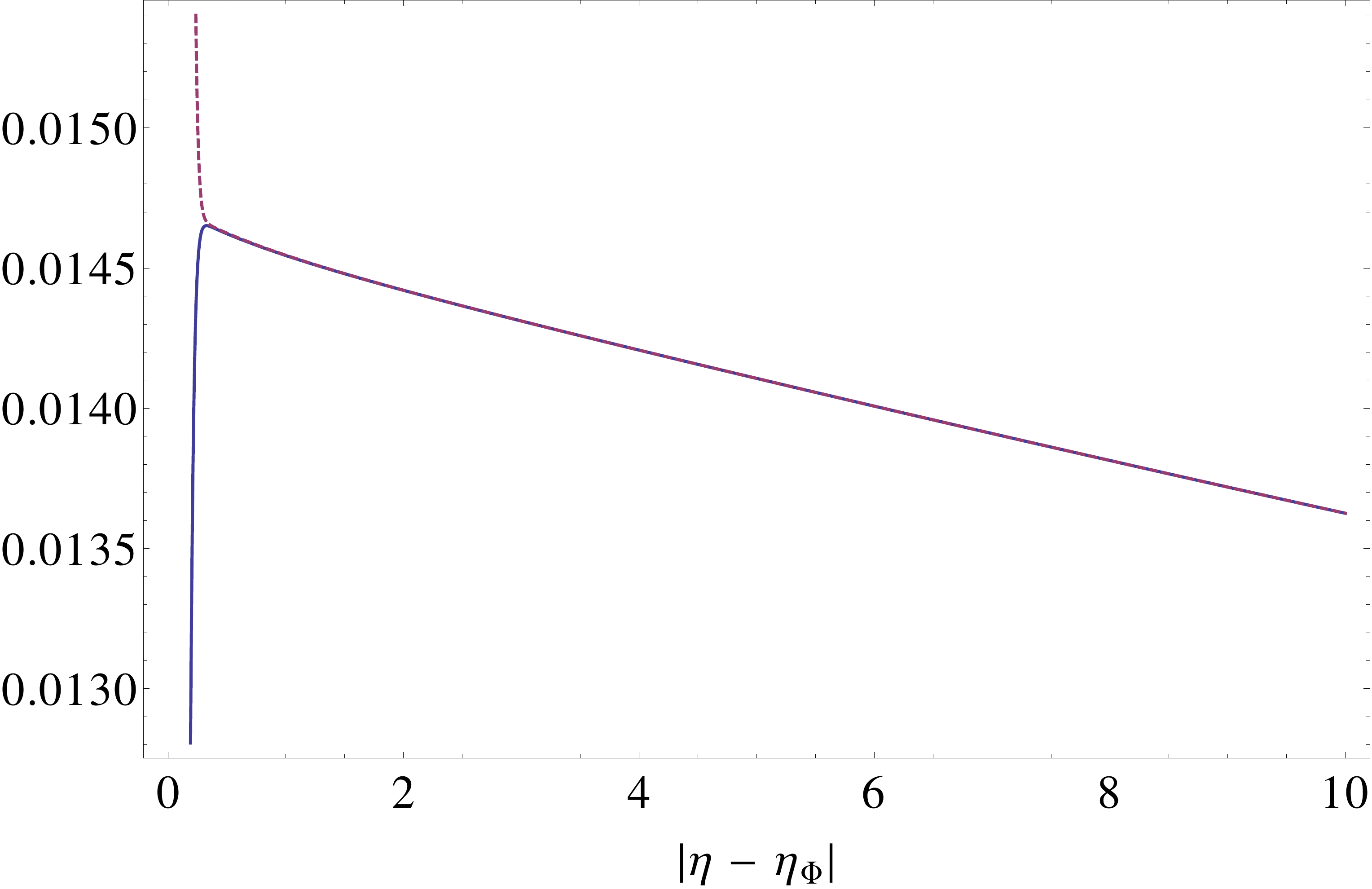}
\caption{This is the plot of $|\mathcal{L}|$ (the solid blue line)
and $\mathcal{H}$ (the dashed purple line) versus
$|\eta- \eta_{\Phi}|$ as we evolve the spacetime backwards in time.
Notice that $|\mathcal{L}|$ converges to $\mathcal{H}$ very quickly.}
\label{attractor}
\end{figure}

For the dominant energy condition to be satisfied we must
have
\be \label{dc}
\rho \geq |p_{\hat{x}_1}|, \hspace{1cm} \rho \geq |p_{\hat{\theta}}|,
\ee
at all times, where $\{ \rho, p_{\hat{x}_1}, p_{\hat{\theta}}\}$ are the
independent eigenvalues of the stress energy tensor in a locally
orthonormal frame
\be 
ds^2 = -d \hat{x}_0 ^2 + d\hat{x}_1 ^2 +d \hat{\theta}^2 + d \hat{\phi}^2
\ee
($p_{\hat{\theta}} = p_{\hat{\phi}}$ due to spherical
symmetry). Given the metric \eqref{metric2}, a routine calculation
gives ($\eta$ dependence is suppressed)
\bea \label{eigenvalues}
&& \rho = \frac{1}{16 \pi a^2}\Big[G_{\eta \eta}-G_{\chi \chi}+\sqrt{(G_{\eta \eta}+G_{\chi \chi})^2-4 G_{\eta \chi}^2}\Big],\nonumber\\
&&p_{\hat{x}_1} = \frac{1}{16 \pi a^2}\Big[-G_{\eta \eta}+G_{\chi \chi}+\sqrt{(G_{\eta \eta}+G_{\chi \chi})^2-4 G_{\eta \chi}^2}\Big],\nonumber\\
&& p_{\hat{\theta}} = \frac{\kappa}{8 \pi a^2 \sin({\sqrt{\kappa} \chi})^2} G_{\theta \theta},
\eea
where $G_{ab}$ is the Einstein's tensor. We should first show that
the quantity inside the square root in Eq. \eqref{eigenvalues} remains
positive semi-definite at all times. This condition is also necessary
to ensure that the eigenvector associated with $\rho$ is timelike. It is sufficient to have 
\bea \label{1ineq}
&& G_{\eta \eta} + G_{\chi \chi} \geq 2 | G_{\eta \chi}| \nonumber\\
&& \Rightarrow 2 \mathcal{H}^2 - 2 \mathcal{H}' + 2 \kappa 
+ \frac{1}{2}\mathcal{L}^2 \big[\chi ^2 \kappa - f(\chi,\kappa) \big]
+ f(\chi,\kappa) \mathcal{L}'\nonumber\\
&& \geq - 2 \chi \kappa \mathcal{L} \nonumber\\
&& \Rightarrow 2 \mathcal{H}^2 - 2 \big[1- g f(\chi,\kappa) \big]\mathcal{H}'
+ 2 \kappa - f(\chi, \kappa) \mathcal{H} \mathcal{L} 
+\frac{1}{2} \chi^2 \kappa \mathcal{L}^2 \nonumber\\
&& \geq - 2 \chi \kappa \mathcal{L},
\eea
where $f(\chi, \kappa) \equiv 1 - \chi \sqrt{\kappa} \cot{(\chi \sqrt{\kappa})}$
and we used the fact that $\mathcal{L}\leq 0$ at all times. We also used Eq. \eqref{flow} in the last inequality. Since $0<\kappa\leq 1$
and $0 \leq \chi \leq \pi/2$, we have $0 \leq f \leq 1 $. Given that
$|\mathcal{L}|< \mathcal{H}$ at all times, the inequality \eqref{1ineq}
comes down to 
\bea
&& 2 \mathcal{H}^2 - 2 \big[1- g f(\chi,\kappa) \big]\mathcal{H}'
+ 2 \kappa\big[1-\frac{\pi}{2} \mathcal{H}\big] - f(\chi, \kappa) \mathcal{H} \mathcal{L} \nonumber\\
&& +\frac{1}{2} \chi^2 \kappa \mathcal{L}^2 \geq 0
\eea
which is correct at times $\eta < \eta_{\Phi}$ because $\mathcal{H} \ll 1$,
$\mathcal{L} \mathcal{H} < 0$, and $[1- g f(\chi,\kappa) \big]\mathcal{H}'<0$,
with the latter following from Eq. \eqref{flow} and $0\leq g < 1$.

To show $\rho \geq |p_{\hat{x}_1}|$, it is sufficient to show 
that $G_{\eta \eta} - G_{\chi \chi} \geq 0$. Using Eq. \eqref{flow},
we have
\bea 
&& G_{\eta \eta} - G_{\chi \chi} = 4 \mathcal{H}^2 + 2[1- g f(\chi,\kappa)]\mathcal{H}' - 3 f(\chi, \kappa) \mathcal{H} \mathcal{L}
+ 4 \kappa\nonumber\\
&&   + \mathcal{L}^2 \Big[1-\frac{1}{2}f(\chi,\kappa)-\frac{3}{2}
\chi \sqrt{\kappa} \cot{(\chi \sqrt{\kappa})} \nonumber\\
&&+\frac{1}{2}\kappa \chi^2
\big(-1+\cot{(\sqrt{\kappa} \chi)}^2 \big) \Big] \geq 4 \mathcal{H}^2 + 2\mathcal{H}'- 3 f(\chi, \kappa) \mathcal{H} \mathcal{L}\nonumber\\
&&+4 \kappa - \mathcal{L}^2 \geq 2 \mathcal{H}^2 + 3 \kappa - \kappa \frac{\mathcal{L}}{\mathcal{H}} - 3 \mathcal{H} \mathcal{L} \geq 0. 
\eea

Finally, we ought to show that $\rho \geq |p_{\hat{\theta}}|$.
As for $ |p_{\hat{\theta}}|$, we have 
\bea \label{ineq31}
&& 8 \pi a^2 |p_{\hat{\theta}}| = \Bigg|-\kappa \frac{\mathcal{L}}{\mathcal{H}}
\bigg[1+\frac{1}{2}g f(\chi,\kappa) \bigg]-\frac{1}{2}g f(\chi,\kappa) [\kappa
+\mathcal{H}^2] \nonumber\\
&&+\frac{1}{2} f(\chi,\kappa) \mathcal{H} \mathcal{L}+\frac{1}{4}\chi^2 \kappa
\mathcal{L}^2 \Bigg| \leq -\frac{3}{2}\kappa \frac{\mathcal{L}}{\mathcal{H}}
+\frac{1}{2}[\kappa + \mathcal{H}^2]-\frac{1}{2}\mathcal{H}\mathcal{L}\nonumber\\
&&+\frac{1}{4} \chi^2 \kappa \mathcal{L}^2 \leq 2 \kappa + \mathcal{H}^2 
\bigg[1+\frac{1}{4} \chi^2 \kappa \bigg] \leq 2[\kappa+\mathcal{H}^2].
\eea
Likewise we have for the locally measured energy density
\bea \label{ineq32}
&& 8 \pi a^2 \rho  = \frac{1}{2} \Big[G_{\eta \eta}-G_{\chi \chi}
+ \sqrt{(G_{\eta \eta}+G_{\chi \chi})^2 - 4 G_{\eta \chi} ^2} \Big]\nonumber\\
&&\geq G_{\eta \eta} - |G_{\eta \chi}| = 3[\kappa + \mathcal{H}^2]
- 2 f(\chi,\kappa) \mathcal{H}\mathcal{L} + \chi \kappa \mathcal{L} \nonumber\\
&&+\frac{1}{2}\mathcal{L}^2 \Big[ f(\chi,\kappa) -\frac{1}{2}
+\frac{1}{2}\kappa \chi^2 \cot{(\chi \sqrt{\kappa})}^2 \Big]\nonumber\\
&& \geq 3[\kappa + \mathcal{H}^2] + \chi \kappa \mathcal{L} \geq
\big[3-\frac{\pi}{2} \mathcal{H}\big] \kappa + 3 \mathcal{H}^2,
\eea
where we used 
\be 
f(\chi,\kappa) -\frac{1}{2}
+\frac{1}{2}\kappa \chi^2 \cot{(\chi \sqrt{\kappa})}^2 >0.
\ee
As can be seen from the last expressions in Eq. \eqref{ineq31} and 
Eq. \eqref{ineq32},for $\rho \geq |p_{\hat{\theta}}|$ it suffices to have
$1-(\pi/2) \mathcal{H} \geq 0$, which is true for $\eta < \eta_{\Phi}$ given
that $\mathcal{H} \ll 1$ at such times. Therefore, we conclude that 
the modified FRW spacetime satisfies the dominant energy condition everywhere
to the past of $\eta_{\Phi}$.

\subsection{A numerical study of the modified Lema\^itre-Tolman-Bondi geometry }
\label{modltb}

We now apply the formalism developed in Sec. \ref{formalism} to the
LTB metric given in Eq. \eqref{LTB}. The equations \eqref{flow} in this case become
\bea \label{flowltb}
&&\dot{R}^2+2 R \ddot{R}+r^2 k \Big(1+ \frac{\dot{k} R}{k \dot{R}}\Big)=0,\nonumber\\
&&\dot{R}\frac{\dot{k}}{k}-\frac{1}{2} R \Big(\frac{\dot{k}}{k}\Big)^2 + \partial_{t} \Big(R \frac{\dot{k}}{k}\Big)=2 g \ddot{R},
\eea
where  dot denotes differentiation with respect to the LTB time coordinate $t$
and $g = 1- \exp{[-\{t_{\Phi}(r)-t\}]}$.  We define $t_{\Phi}(r)$ to be the
location of the spacelike slice  between $t_{\Sigma}$ and the future
event horizon found in Sec. \ref{m2} for which all spherical shells have
\bea \label{rdot}
&&\dot{R}[t_{\Phi}(r),r]\approx \frac{\Big(\frac{\pi}{2}-3k(r)^{3/2} [t_{\Phi}(r)-t_0 (r))\Big]}{2} r \sqrt{k(r)} \nonumber\\
&&= - \frac{10^{-6}}{2} r \sqrt{k(r)},
\eea
where $k(r)$ is the spatial curvature that we constructed in Sec. \ref{m2}.
The relation given in the first line of Eq.\eqref{rdot} comes from
the Taylor expansion of $\dot{R}$ near $t_{\Sigma}$ for which $\dot{R}=0$
everywhere. The numerical factor $10^{-6}$ in the second line of
Eq. \eqref{rdot} is chosen so that all spherical shells are located
between $t_{\Sigma}$ and the future event horizon.

We numerically solve the coupled system of ordinary differential equations
\eqref{flowltb} for $-\infty< t < t_{\Phi}(r)$ and $0 \leq r \leq 99/10$ using the following initial conditions \footnote{Recall that we are integrating backwards in time, so
$\dot{R}>0$.}
\bea
&&k[t_{\Phi}(r),r]=k(r),\hspace{1 cm}\dot{k}[t_{\Phi}(r),r]=0,\nonumber\\
&&\dot{R}[t_{\Phi}(r),r]=\frac{10^{-6}}{2} r \sqrt{k(r)}.
\eea
We then examine the locally measured energy density and principal 
pressures of the modified spacetime, which are given by 
\bea
&& \rho = \frac{1}{16 \pi}\Big[G_{t t}-G^{r}_{r}+\sqrt{(G_{tt}+G^{r}_{r})^2 - 4 G_{tr} G^{r}_{t}}\Big],\nonumber\\
&&p_{\hat{x}_1} = \frac{1}{16 \pi}\Big[-G_{t t}+G^{r}_{r}+\sqrt{(G_{tt}+G^{r}_{r})^2 - 4 G_{tr} G^{r}_{t}}\Big],\nonumber\\
&&p_{\hat{\theta}} = \frac{1}{8 \pi}G^{\theta}_{\theta}.
\eea

\begin{figure}
\centering
\includegraphics[width=0.45 \textwidth]{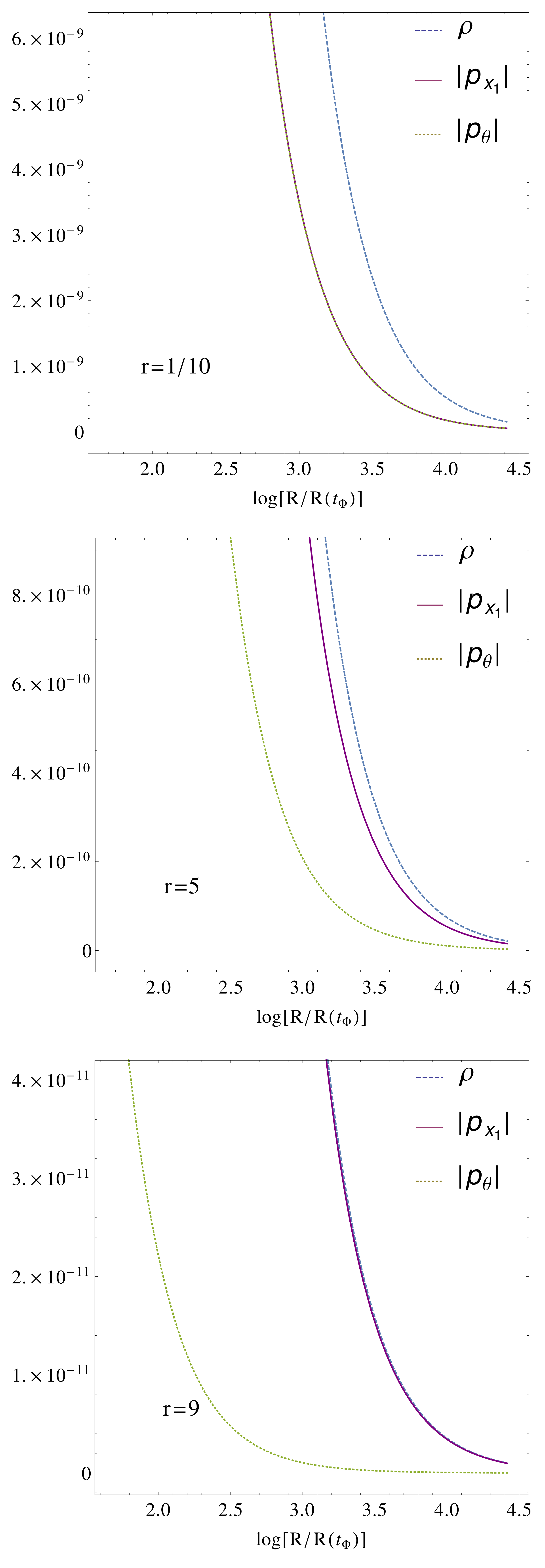}
\caption{This is the plot of the eigenvalues of the
modified stress energy tensor versus $\log[R(t,r)/R(t_{\Phi},r)]$
for spherical shells at three different radii.}
\label{ltbdominant}
\end{figure}

The modified LTB spacetime shares very similar qualitative features
to its homogeneous counterpart. In particular, $\dot{k}/k
\rightarrow - \dot{R}/R$, or equivalently $R \propto t^{2/3}$
and $k \propto t^{-2/3}$,  at times $t \ll t_{\Phi}(r)$ for all
spherical shells. Similar to the case of the modified FRW metric,
it is not difficult to check that the constant $t$ slices
become nearly intrinsically flat as $t \rightarrow -\infty$.
Despite the fact that the eigenvalues
of the modified stress energy tensor no longer have the same
functional form as the ones obtained in the previous section as
they now contain terms with radial derivatives, our extensive numerical
investigations provide no evidence for the violation of the
dominant energy condition. Therefore, we are convinced that
the modified LTB spacetime constructed using \eqref{flowltb}
avoids the white hole singularity while satisfying the dominant
energy condition. In Fig. \ref{ltbdominant} we plot the eigenvalues
of the modified stress energy tensor in a local orthonormal frame
for spherical shells at a few different radii.

Finally, note that there is also a non-vanishing boundary stress energy
tensor for $t<t_{\Phi}(r_{\text{out}})$.
This is due to the discontinuity of the time-time component of the boundary extrinsic curvature in this region of spacetime.
Indeed, $\vec{u} = \vec{\partial_{t}}$ is no longer a geodesic in the exterior Schwarzschild geometry for $t<t_{\Phi}(r_{\text{out}})$, while remaining a geodesic in the interior geometry at all times. The independent eigenvalues of the boundary stress energy tensor in the orthonormal $\{t,\hat{\theta},\hat{\phi}\}$ frame are given by \footnote{The boundary stress energy tensor comes from a standard calculation for the exterior Schwarzschild geometry. We have
also used Eq. \eqref{flowltb}. See \citep{poisson}.}
\bea
&& \bar{\rho} = 0, \hspace{1cm} \bar{p} = -\frac{1}{16 \pi}
\frac{r^2 k }{R \sqrt{1-\frac{2M}{R}+\dot{R} ^2}} \frac{\dot{k} R}{k \dot{R}},
\eea
which are evidently positive semi-definite at all times due to the fact that $(\dot{k} R)/(k \dot{R}) \leq 0$ [See Fig. \ref{attractor}]. Therefore, the boundary stress energy tensor satisfies the weak energy condition.

\subsection{Black hole's formation timescale}
\label{time}

\begin{figure}[h]
\centering
\includegraphics[width=0.5 \textwidth]{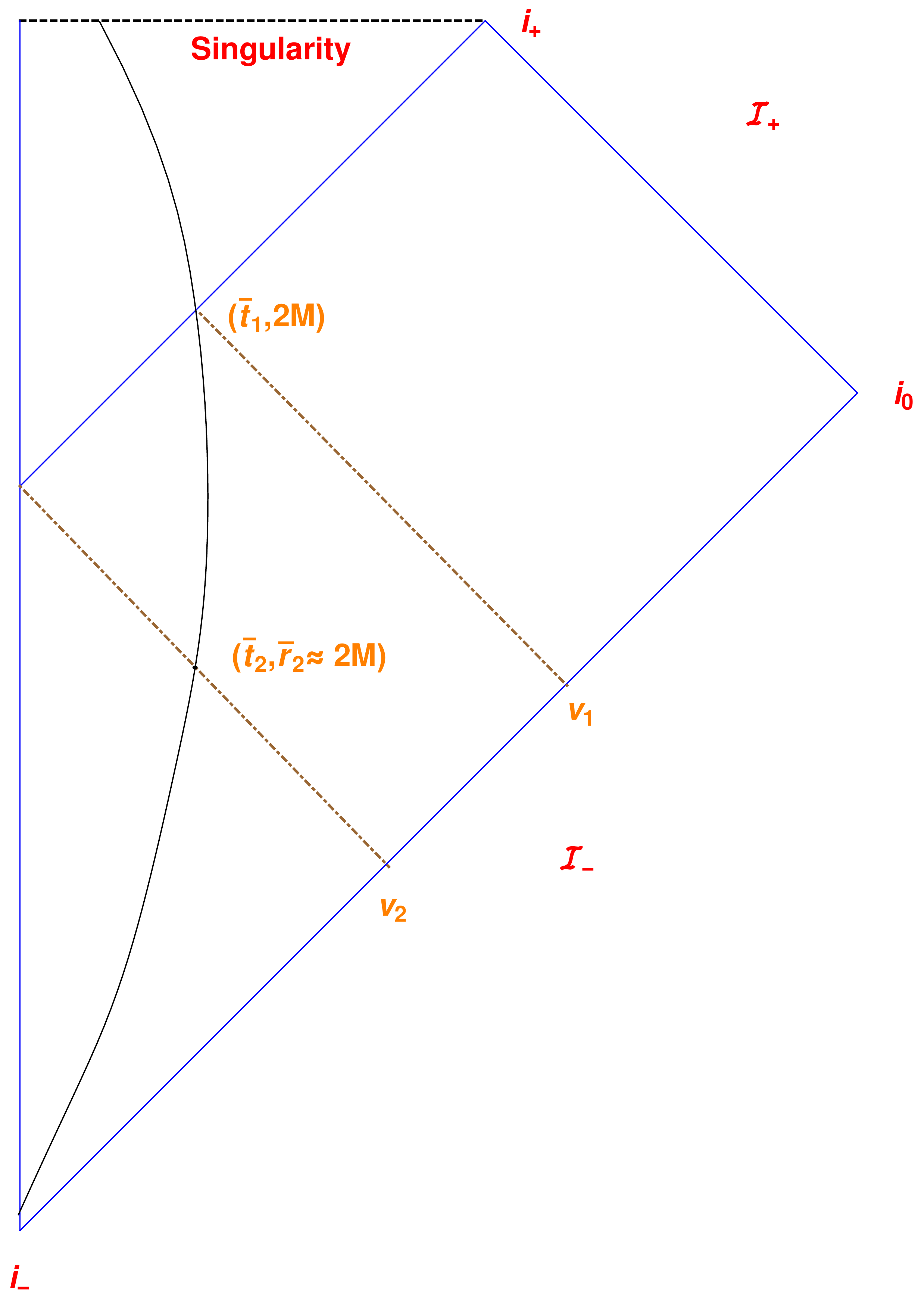}
\caption{This is the Penrose diagram of a collapsing object
embedded in a Schwarzschild exterior. The ingoing null rays with
advanced Eddington-Finkelstein coordinates $v_1$ and
$v_2$ intersect the collapsing object at $(\bar{t}_1,2M)$ and
$(\bar{t}_2, \bar{r}_2 \approx 2M)$ respectively.}
\label{penrose}
\end{figure}

We find it appropriate to define the black hole's formation
timescale to be the interval along $\mathcal{I}_{-}$
between the two ingoing null geodesics that intersect
the beginning of the event horizon and the outer boundary of the collapsing
object as it crosses the event horizon [See Fig. \ref{penrose}].
Intuitively, this corresponds to the time
required for the collapsing object to cross its own event horizon
as measured by the asymptotic null observers at $\mathcal{I}_{-}$.

We now show that this timescale is on the order of $M^2$ for the modified
LTB collapse models constructed in Sec. \ref{modltb}. To begin, notice that
the modified LTB geometry evolves very slowly compared to the LTB geometry.
In fact, to a good approximation we have
\be
\Delta v = v_1 - v_2 = \Delta \bar{t} + \Delta \bar{r}_{*} \approx \Delta \bar{t},
\ee
where $\bar{r}_{*}$ is the Schwarzschild's tortoise coordinate \footnote{$\bar{r}_{*} \equiv \bar{r} + 2 M \log{|\bar{r}/(2M) -1|}$}.
In other words, the
proper observer on the boundary is nearly stationary between the
times $\bar{t}_1$ and $\bar{t}_2$. As a result, we have
\be \label{timescale1}
\Delta \bar{t} \approx \frac{\Delta t}{\sqrt{1-\frac{2M}{R(t_1,r_{\text{out}})}}},
\ee
where $\Delta t$ is the proper time interval between $\bar{t}_1$ and $\bar{t}_2$
measured by the proper observer on the boundary. Using the fact that the proper observer
on the boundary remains quite close to its initial location where $\dot{R}\approx0$, it follows from the
Eqs. \eqref{bang}, and \eqref{saturated} that
\be \label{timescale2}
\Delta \bar{t} \sim  M \rho_{\text{out}} ^{1/4} \Delta t \sim M \rho_{\text{out}} ^{1/2} S
\sim M^3 \rho_{\text{out}} ^{1/2} \sim M^2,
\ee
where $S$ is the entropy evaluated on $\Sigma$ and $\rho_{\text{out}}\equiv \rho[t_{\Phi}(r_{\text{out}}),r_{\text{out}}] \sim r_{\text{out}}^{-6} \sim M^{-2}$, assuming that $k \sim r^{-2}$ for the outer
shells [See Sec. \ref{m2}].
To arrive at \eqref{timescale2}, we also used the slow evolution
of the interior metric to argue that $\Delta t$ is on the order of the
temporal change along the future event horizon. We then approximated
the temporal change along the event horizon using its value along
the surface $\Sigma$. This is a reasonable approximation
given the proximity of $\Sigma$ to the event horizon.

\section{concluding remarks}

In this paper we revisited the validity of  
the conjectured entropy-mass-timescale relation given in \eqref{genericrel}.
We constructed a pathology free spacetime from a monster-like initial condition that saturates the Bekenstein-Hawking entropy bound and forms a black hole within a timescale on the order of $M^2$. The constructed spacetime satisfies 
the weak energy condition. This construction invalidates the conjectured
entropy-mass-timescale relation by a factor of $M^{1/4}$.

However, the constructed spacetime appears to be finely tuned 
and probably does not represent a generic scenario of gravitational collapse. Moreover, though beyond the scope of this paper, the question of whether the spacetimes constructed
using the recipe provided in Eqs. \eqref{flow} and \eqref{flowltb}
are classically and quantum mechanically stable needs to be addressed.
Perhaps one way of looking at our conclusion is that violating 
the conjectured entropy-mass-timescale relation can only be done
by some contrived mathematical construction.

Finally, we note that discovering any connection between the relation given in Eq. \eqref{genericrel} and the quantum focusing
conjecture formulated in \citep{boos} can be illuminating.

\section{acknowledgments}

We are very grateful to \'Eanna  Flanagan for his comments on the manuscript.
We also thank Leonard Gross, Nils Deppe, and Alexander Grant for helpful discussions.

\appendix

\section{ Lema\^itre-Tolman-Bondi geometry with event horizon intersecting the singularity}\label{nocauchy}
\begin{figure}[h]
\includegraphics[width=0.5\textwidth]{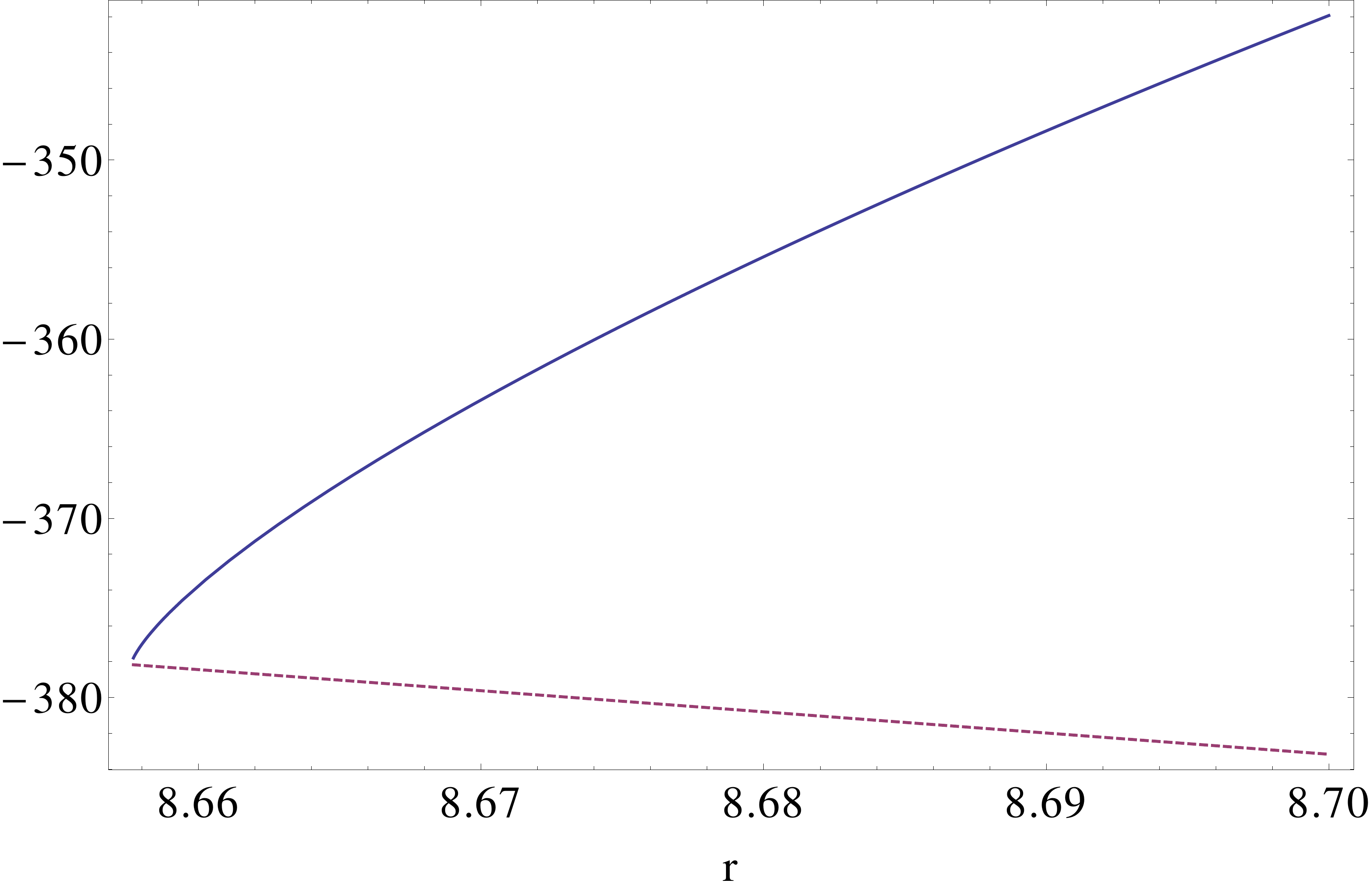}
\caption{This is the plot of $t_{\text{H}}$ (solid blue line)
and $t_0(r)$ (dashed violet line) versus $r$. The future event horizon intersects the curvature
singularity at $r \approx 8.657$.}
\label{eve-sing}
\end{figure}

In Sec. \ref{m2} we mentioned that it is possible for
some LTB-Schwarzschild geometries not to have any
achronal slices that are completely outside of the future event horizon.
It is then implied that in these spacetimes the future event horizon intersects the past curvature singularity.
As an example, let us assume that the interior LTB geometry has
a radial coordinate $r$ that covers from $r=0$ to $r=99/10 \equiv r_{\text{out}}$.
Now consider a constant time slice $\Theta$ for which we set $t=0$ and that it is foliated by the maximal spheres, i.e. $\dot{R}(t_{\Theta},r)=0$ everywhere on $\Theta$. After setting $8 \pi \bar{\rho} = 1$, it follows from  Eq. \eqref{sol}  that $u= \pi$ and
\be
t_0 (r) = - \frac{\pi}{6 k(r)^{3/2}}.
\ee
As for $k(r)$, we take the same solution as in Sec. \ref{m2} by assuming
that the non-degeneracy condition \eqref{validity} the shell-apparent horizon
physical distance condition \eqref{shell-apphori} are both saturated.
We then solve Eq. \eqref{event} to find the location of the future event
horizon subject to the appropriate boundary value for $r_{\text{out}}$.
It turns out that the future event horizon intersects the past curvature
singularity at $r \approx 8.657$. See Fig. \ref{eve-sing}

\section{On the validity of the Gaussian normal coordinates for the modified metrics}

\begin{figure}[h]
\includegraphics[width=0.5\textwidth]{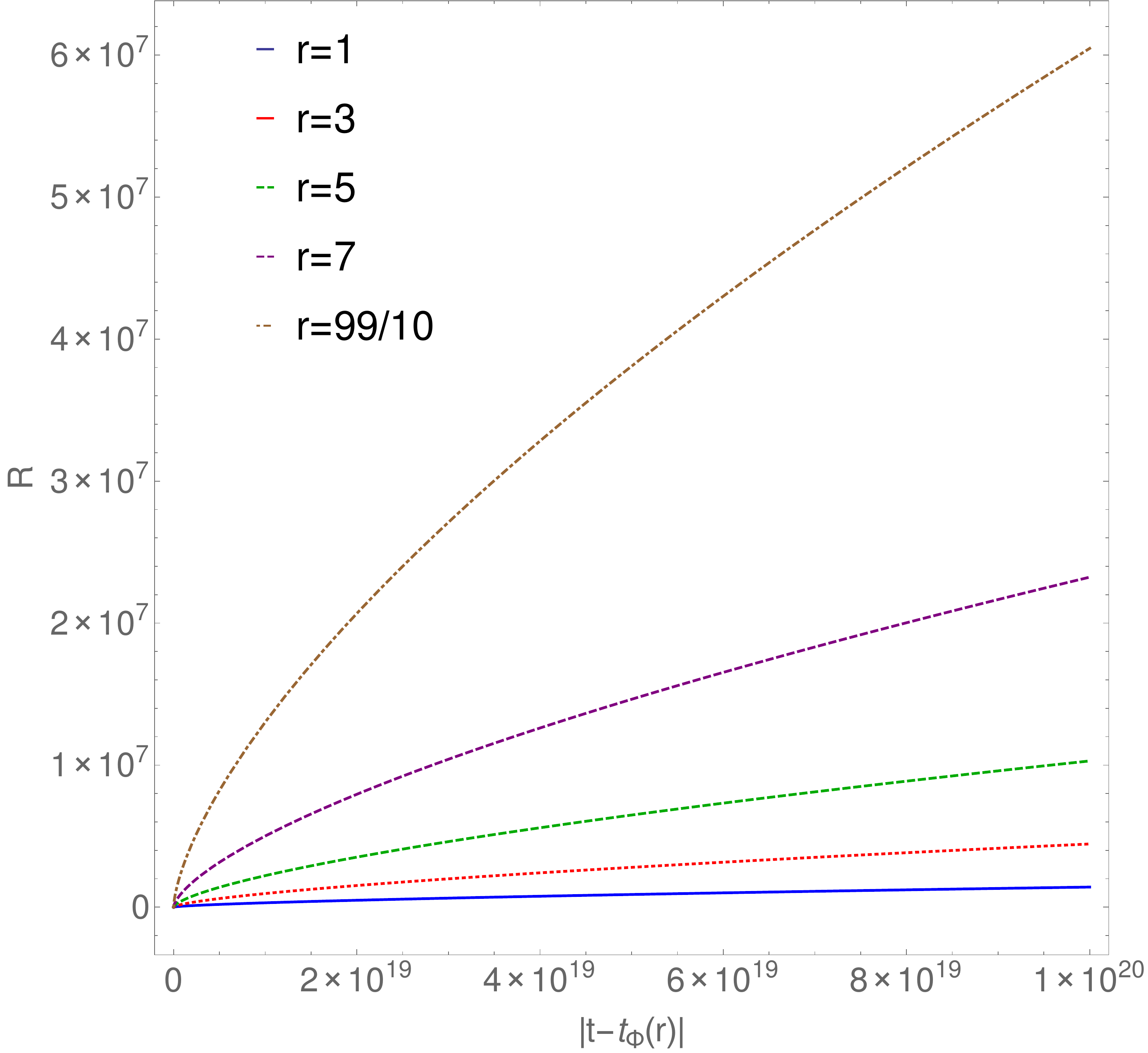}
\caption{Here we have plotted the solution of Eq. \eqref{flowltb}
for a number of shells. As can be seen from the figure,
shells expand without any crossing.}
\label{shell-not-crossing}
\end{figure}

In Sec. \ref{circ} we introduced a formalism in which the
radial-radial component of the FRW or LTB metric
had an additional time dependence. We implicitly assumed
that the Gaussian normal coordinates used to define the
FRW and LTB metrics can also be used to define the modified
metrics. Here we sketch an argument showing the validity
of this assumption.

Recall that the Gaussian normal coordinates can be constructed
for a spacetime if and only if there exists a family of $C^{n\geq1}$ timelike
geodesics in the spacetime for which every point is located
on exactly one of these timelike geodesics. The coordinates
are then constructed with $\vec{u} = \vec{\partial}_{t}$ being the
unit tangent to each timelike geodesic. A trivial consequence of the
existence of such a family of timelike geodesics is that
there are no congruences of timelike geodesics with unit
tangent $\vec{u} = \vec{\partial}_{t}$ that form a caustic
anywhere in the spacetime \footnote{Except at the curvature singularity, which
is formally not  part of the spacetime.}.

The modified LTB metric is given by
\be
ds^2 = - dt^2 + \frac{R'(t,r)^2 dr^2}{1-r^2 k(t,r)} + R(t,r)^2 d \Omega^2,
\ee
where $r^2 k(t,r)<1$ at all times and $\dot{k}(t,r)\neq 0$ for $-\infty<t<t_{\Phi}(r)$
[see Sec. \ref{modltb}]. Here we have implicitly assumed that the
Gaussian normal coordinates used to define the LTB metric \eqref{LTB}
can be used to define the modified LTB metric as well. To ensure
that this can be done, we must show that the expansion
$\theta \equiv {}^{(3)}g^{ab} \nabla_{a} u_{b}$ does not diverge
for any congruence of timelike geodesics with
unit tangent $\vec{u}=\vec{\partial}_{t}$ \footnote{${}^{(3)}g^{ab}$
is the inverse metric for constant $t$ slices.}.
A short calculation gives
\bea \label{expansion}
&&\theta = \frac{1}{2} {}^{(3)}g^{ab} \dot{g}_{ab} \nonumber\\
&& = \frac{\dot{R}^{\prime}(t,r)}{R^{\prime}(t,r)}+ \frac{\dot{R}(t,r)}{R(t,r)}+\frac{r^2 \dot{k}(t,r)}{2 [1-r^2 k(t,r)]}.
\eea
Pursuant to our discussions in Sec. \ref{formalism} and \ref{modltb} with
regards to the modified metric functions and their asymptotic behaviour, we
do not expect any divergences to occur in \eqref{expansion} at any radii
and time in the past. Therefore, we do not see a problem
with the use of Gaussian normal coordinates for the modified FRW and
LTB metrics.

\section{Aerial radii do not coincide for distinct coordinate radii in the modified spacetime}

If two aerial radii corresponding to two distinct 
coordinate radii coincide at some
time to the past of $t_{\Phi}(r)$,
the formalism of Sec. \ref{circ} would be invalidated.
Numerically we can confirm that this does not happen 
for the solutions to the system of equations \eqref{flowltb}.
See Fig. \ref{shell-not-crossing}

\bibliographystyle{unsrtnat}
\bibliography{Reference}
\end{document}